\PassOptionsToPackage{unicode}{hyperref}
\PassOptionsToPackage{hyphens}{url}
\documentclass[
  11pt]{article}
\usepackage{xcolor}
\usepackage[margin=1in]{geometry}
\usepackage{amsmath,amssymb}
\usepackage{iftex}
\ifPDFTeX
  \usepackage[T1]{fontenc}
  \usepackage[utf8]{inputenc}
  \usepackage{textcomp} 
\else 
  \usepackage{unicode-math} 
  \defaultfontfeatures{Scale=MatchLowercase}
  \defaultfontfeatures[\rmfamily]{Ligatures=TeX,Scale=1}
\fi
\usepackage{lmodern}
\ifPDFTeX\else
\fi
\IfFileExists{upquote.sty}{\usepackage{upquote}}{}
\IfFileExists{microtype.sty}{
  \usepackage[]{microtype}
  \UseMicrotypeSet[protrusion]{basicmath} 
}{}
\makeatletter
\@ifundefined{KOMAClassName}{
  \IfFileExists{parskip.sty}{%
    \usepackage{parskip}
  }{
    \setlength{\parindent}{0pt}
    \setlength{\parskip}{6pt plus 2pt minus 1pt}}
}{
  \KOMAoptions{parskip=half}}
\makeatother
\usepackage{longtable,booktabs,array}
\usepackage{caption}
\usepackage{calc} 
\usepackage{etoolbox}
\makeatletter
\patchcmd\longtable{\par}{\if@noskipsec\mbox{}\fi\par}{}{}
\makeatother
\IfFileExists{footnotehyper.sty}{\usepackage{footnotehyper}}{\usepackage{footnote}}
\makesavenoteenv{longtable}
\usepackage[]{natbib}
\usepackage{amsmath}
\usepackage{booktabs}
\usepackage{longtable}
\usepackage{array}
\usepackage{graphicx}
\usepackage{hyperref}
\providecommand{\xmpquote}[1]{#1}
\usepackage{bookmark}
\IfFileExists{xurl.sty}{\usepackage{xurl}}{} 
\makeatletter
\@ifundefined{xmpquote}{\newcommand{\xmpquote}[1]{#1}}{}
\makeatother
\hypersetup{
  pdftitle={LLM-Assisted Review Prioritization for German Statutory Health Insurance Websites: A Multi-Stage Corpus Audit},
  pdfauthor={Martin Möller},
  pdfkeywords={\xmpquote{public health communication}, \xmpquote{health
information quality}, \xmpquote{large language
models}, \xmpquote{web-scale audit}, \xmpquote{review
prioritization}, \xmpquote{regulatory informatics}, \xmpquote{evidence
governance}, \xmpquote{statutory health
insurance}, \xmpquote{Germany}, \xmpquote{AI-generated
content}, \xmpquote{inter-model agreement}, \xmpquote{public
observability}},
  hidelinks,
  pdfcreator={LaTeX via pandoc}}

\title{LLM-Assisted Review Prioritization for German Statutory Health
Insurance Websites: A Multi-Stage Corpus Audit}
\author{Martin Möller}
\date{2026-07-31}

\begin{document}
\maketitle
\begin{abstract}
Background: German statutory health insurance (SHI) funds publish web
portfolios that exceed continuous specialist review capacity. Their
content can shape health and benefit expectations. Generic AI-text
detection does not identify medical, benefit, legal, or editorial review
needs. Objective: To characterize a multi-stage workflow that
prioritizes substantive review needs while separating AI-provenance
signals from quality claims. Methods: We analyzed 56,198 pages from 84
SHI websites or sub-sites. The workflow combined deterministic
screening, model-assisted triage and in-depth review, minimum evidence
checks, temporal-validity safeguards, and paired-model comparison. It is
reproducibility-bounded, not a validated detector. Production code is
proprietary; reproducibility rests on frozen derived tables and
paired-comparison artifacts. Public-observability coding was exploratory
and model-assisted. The 300-page lower-priority check was a
single-model, risk-enriched routing stress test, not a human-reference
evaluation. Results: All pages received a review state. The workflow
generated 35,998 review records and routed 21,452 to case review. The
workload concentrated in transparency, legal framing, medical content,
contradictions, and AI-related failure-mode signals. Of 33,448
concept-eligible records, 23,273 linked to a concept; 290 patterns held
21,459 assignments over 17,076 distinct records, since a record could
appear under several patterns, and 6,197 concept-linked records entered
no pattern. A quoted passage was locatable in captured page text for
31,347 records, confirming literal occurrence rather than factual
correctness. The routing stress test surfaced a signal on 100/300 pages
(33.3\% within the sample). Across 182 matched cases, two models agreed
in 75.8\% (kappa = 0.532; 95\% CI 0.415-0.649). Exploratory coding
suggested that several governance signals were not publicly
discoverable. Conclusions: The workflow produces a prioritized workload,
not error prevalence or final legal, medical, or insurer-level findings.
It neither proves AI authorship nor validates autonomous detection.
Paired-model agreement quantifies consistency, not correctness or
sufficient triage performance; public claims require human adjudication.
\end{abstract}

\section{Introduction}\label{introduction}

Public websites of German statutory health insurance (SHI) funds
(gesetzliche Krankenkassen) form a high-volume layer of public health
communication. Specialist editorial, medical, legal, and benefit teams
cannot continuously recheck every page manually. The methodological
problem is evidence-based review prioritization under uncertainty: how
to turn a large public web corpus into a reproducibility-bounded,
prioritized set of passages that merit review without treating automated
review as final judgment.

These websites do not replace individual medical care, legal advice, or
formal benefit decisions, but they shape insured persons' understanding
of prevention, disease, treatment options, statutory benefits, and
procedural rights. Under German social law, SGB V assigns sickness funds
a role in helping insured persons through information, counselling, and
benefits (§ 1 SGB V), and the quality and effectiveness of benefits must
reflect the generally accepted state of medical knowledge (§ 2 Abs. 1
SGB V). Sections 13--14 SGB I address information and counselling, while
§ 305 Abs. 3 SGB V requires comprehensive information on approved
providers, prescribable services, and sources of supply upon request; it
is a request-based information duty, not a public publication duty
\citep{sgbv1,sgbv2,sgbi13,sgbi14,sgbv305}. Inaccurate, outdated,
insufficiently conditioned, or evidentially unsupported content is
therefore an editorial-quality problem with potential implications for
health literacy and benefit expectations.

Public-facing health content is now produced in an environment where
generative language models can draft, adapt, translate, and repurpose
large volumes of web content at low marginal cost. Web-scale analyses
have reported a substantial increase in AI-generated or AI-assisted web
pages after the public release of large conversational models
\citep{dolezal2026}. For regulated health communication, the central
risk is not AI-assisted authorship by itself. The relevant question is
whether content-production velocity, model-mediated rewriting, and
inherited archive material outpace the editorial, medical, and legal
controls required for patient-facing public information. A human-written
page can be materially problematic; an AI-assisted page can be
substantively correct if it has undergone competent review. In this
study, provenance is treated as a probabilistic and explanatory signal,
not as a substitute for substantive evaluation.

This regulatory tension is visible at the boundary between public health
information, health-related advertising, and AI transparency. Where a
statement constitutes advertising within the meaning of § 1 HWG, § 3 HWG
can provide a standard for assessing misleading health-related
advertising \citep{hwg3}. The corresponding audit pattern marks a need
to assess applicability and evidentiary support, not a violation.
Article 50(2) of the EU Artificial Intelligence Act addresses providers
of systems that generate synthetic content; Article 50(4) addresses
deployers that publish AI-generated or manipulated text on matters of
public interest, and its exception concerns human review or editorial
control under editorial responsibility \citep{euaiact2024}. The European
Commission published its final transparency guidelines on 20 July 2026
\citep{eucomguidelines2026}. The May 2026 data collection preceded the
obligations' applicability from 2 August 2026 and was not a compliance
assessment. Neither AI-style patterns nor the absence of a visible AI
label establishes a legal conclusion for an SHI page.

Used alone, current approaches do not solve this task. Generic AI-text
detection estimates stylistic provenance but does not determine whether
a statement is medically current, legally conditioned, or supported by
authoritative sources. Single-pass large-language-model review can
identify substantive issues but is sensitive to calibration, prompt
framing, page length, and model knowledge cutoffs. Manual specialist
review remains necessary for high-stakes adjudication, yet it does not
scale to tens of thousands of heterogeneous web pages across insurers,
archives, magazines, service sections, press releases, and benefit
descriptions. The operational task is to make review prioritization
auditable at the level reported in this manuscript: transforming a large
public web corpus into a prioritized set of evidence-linked statements
requiring medical, legal, or editorial review.

The study was organized around three descriptive research questions.
RQ1: What substantive review workload does the audit surface once review
records must retain page-level evidence, materiality fields, and basic
page-evidence checks? RQ2: Can the workflow assign a current review
state to a complete production corpus of public SHI web pages while
preserving the distinction between broad triage and in-depth review?
RQ3: Which calibration limits become visible when a subset of flagged
review cases is assessed by two audit model configurations and
interpreted through paired agreement statistics rather than single-model
authority? These questions keep review workload, page-level coverage,
and case-level reliability evaluation separate. They do not produce
page-level legal determinations about named insurers or treat AI-style
provenance signals as substantive quality findings.

The audit architecture was designed for this problem. It separates
provenance-oriented signals (P) from substantive quality signals (Q), so
that stylistic indicators of AI-assisted production do not become the
evidentiary basis for quality or compliance claims. It combines
deterministic crawling and screening, page-level scoring, lower-cost
triage, domain-grounded in-depth review, temporal-validity safeguards
for post-cutoff legal and medical change, and human-in-the-loop
adjudication of review records. The present manuscript reports a
corpus-scale empirical implementation of this architecture for German
SHI web portals: n = 56,198 publicly accessible pages from 84 website
entities (websites, portals, or sub-sites; defined in Methods), with all
pages assigned a current review state in the corpus coverage snapshot.
The empirical aim is to assess whether a scalable compliance-audit
pipeline can surface, prioritize, and document candidate review needs in
public health-insurance information while keeping model-driven
probability, page-level evidence, and human adjudication distinct.

The contribution is implementation-oriented rather than
model-architectural. This study evaluates the operational feasibility of
scaling compliance audits for public health-insurance websites under
real regulatory and evidentiary conditions, including the German
social-law context of SGB V, health-advertising law, the EU Artificial
Intelligence Act, changing public web content, heterogeneous source
access, and the continuing need for specialist adjudication.

This paper contributes (1) a corpus-scale review workload outside
synthetic benchmarks; (2) P/Q decoupling of provenance signals from
substantive review priorities; (3) a temporal-validity safeguard for
shared page-and-model knowledge cutoffs; (4) a dual-model comparison
that quantifies consistency and divergence workload; (5) exploratory,
model-assisted, and not independently human-verified observability
coding indicating limited discoverability under the public search path;
and (6) an implementation observation that automated cross-source
comparison remains partial and source-dependent because authoritative
sources differ in interfaces and versioning. Together these
contributions instantiate implementation-oriented compliance auditing
for public digital health communication \citep{knott2026}.

\subsection{Reader Map}\label{reader-map}

This is a report on review prioritization, not a catalogue of confirmed
errors. The table is the canonical claim and denominator boundary for
the manuscript.

{\def\LTcaptype{none} 
\begin{longtable}[]{@{}
  >{\raggedright\arraybackslash}p{(\linewidth - 6\tabcolsep) * \real{0.2308}}
  >{\raggedleft\arraybackslash}p{(\linewidth - 6\tabcolsep) * \real{0.3077}}
  >{\raggedright\arraybackslash}p{(\linewidth - 6\tabcolsep) * \real{0.2308}}
  >{\raggedright\arraybackslash}p{(\linewidth - 6\tabcolsep) * \real{0.2308}}@{}}
\toprule\noalign{}
\begin{minipage}[b]{\linewidth}\raggedright
Analytic level
\end{minipage} & \begin{minipage}[b]{\linewidth}\raggedleft
Denominator or output
\end{minipage} & \begin{minipage}[b]{\linewidth}\raggedright
Supported claim
\end{minipage} & \begin{minipage}[b]{\linewidth}\raggedright
Unsupported claim
\end{minipage} \\
\midrule\noalign{}
\endhead
\bottomrule\noalign{}
\endlastfoot
Corpus coverage & 56,198 pages & Every page received a recorded state &
Equal review depth \\
Review workload & 35,998 records; 21,452 routed to case review & Stored
passages require further assessment and define capacity demand & Error
prevalence or confirmed findings \\
Evidence status & 31,347 located; 4,631 not located; 20 missing &
Literal quote occurrence in captured page text where located & Factual
correctness or source validity \\
Concept and pattern layer & 33,448 eligible; 23,273 linked; 21,459
assignments in 290 patterns & Non-disjoint cross-entity prioritization &
A funnel, distinct case count, legal violation, or insurer ranking \\
Routing stress test & 300 enriched pages & Single-model residual signal
within the sample & Sensitivity or full-corpus missed-routing
prevalence \\
Paired-model comparison & 182 matched cases & Consistency and divergence
workload & Correctness or sufficient triage performance \\
Public observability & 420 cells across 84 entities and 5 signals &
Discoverability under the defined search path & Internal governance
practice or effectiveness \\
Public case & Human-adjudicated case & Release-grade claim after quote,
source, context, and counterargument review & Publication of
unadjudicated model output \\
\end{longtable}
}

Provenance-oriented (P) signals can explain or route possible
AI-assisted production, whereas quality-oriented (Q) signals identify
content for substantive review. Neither signal type bypasses the
public-case requirements.

\section{Related Work}\label{related-work}

This study draws on five adjacent literatures. First, European
healthcare quality-management work frames LLMs as tools for
administrative review, compliance monitoring, and quality assurance
rather than as direct patient-care systems \citep{knott2026}. The
present workflow follows that boundary: it does not generate health
advice or decide benefits; it audits already published public content
and produces review priorities for human adjudication.

Second, medical-LLM evaluation work cautions against single-score
validation. Chen et al.~show that medical LLM benchmarks often
under-specify clinical fidelity, robustness, data management, and
governance \citep{chen2025}. Reviews of healthcare LLM applications
reach a similar conclusion: much evaluation remains accuracy-centered,
benchmark-heavy, and weak on calibration, deployment context, and
robustness \citep{bedi2025,busch2025}. For the present study, the lesson
is methodological. Public SHI pages are production materials that
insured persons may actually read, but model output over those pages
still has to be treated as a review signal rather than as final truth.

Third, LLM-as-judge work helps interpret agreement without overreading
it. Zheng et al.~report high agreement between strong LLM judges and
human preferences in assistant-response evaluation \citep{zheng2023}.
That setting is not legal, medical, or editorial adjudication of
health-insurance content. It is useful only as a cautionary comparator:
raw agreement can support routing, while chance-corrected agreement and
specialist review remain necessary for public claims.

Fourth, web-scale studies of AI-generated text establish the background
condition that large volumes of web content may now be drafted,
rewritten, or assembled with model assistance \citep{dolezal2026}. A
general web-wide estimate does not answer the domain-specific question
here. In SHI communication, the relevant risk is not AI authorship as
such, but whether published benefit, medical, legal, and editorial
statements remain current, sufficiently conditioned, and reviewable
against authoritative sources.

Fifth, domain-specific surveys of German statutory health insurance web
content provide the closest empirical precedent. Scherenberg and Preuss
conducted a manual inventory of digital-health-literacy offerings across
97 SHI funds and reported limited central discoverability and fragmented
provision \citep{scherenberg2023}. Greater provider, editorial, and
funding transparency was an author recommendation, not an independently
measured prevalence finding. The discoverability result converges with
the present public-observability module; manual scope and a descriptive
point-in-time inventory remain shared constraints. Scherenberg, Mueller,
and Erhart provide a complementary demand-side perspective by reviewing
public AI language models as consumer health-information sources
\citep{scherenberg2025}. They describe mostly correct but generic and
incomplete answers, continued higher trust in physicians, and a risk
that unguided use widens health-literacy inequalities. Their study
treats AI as a source used directly by insured persons, whereas the
present study uses AI as an audit instrument over insurer-published
content. Both address the quality and reviewability of public health
information from different ends of the information chain.

The gap is operational. Existing work explains why LLM review must be
governed and calibrated; it does not show how a large health-insurance
web corpus can become a prioritized, evidence-preserving review workload
while keeping provenance indicators separate from substantive quality
review.

\section{Methods}\label{methods}

\subsection{Study Design And Corpus}\label{study-design-and-corpus}

This study is a retrospective, observational audit of publicly
accessible web pages published by German statutory health insurance
website entities. The unit of analysis is the individual public web page
after crawl extraction and Markdown normalization. No private
insured-person data, claims data, patient records, or user-interaction
logs are part of the corpus. The corpus coverage snapshot (2026-05-19)
comprises n = 56,198 pages from 84 website entities. The corpus is not a
random sample; it is a production website-portfolio audit designed to
maximize coverage of publicly available insurer information.

For terminology, \texttt{SHI} refers to German statutory health
insurance. \texttt{GKV} is retained only where the German label itself
is relevant, such as \texttt{GKV-Spitzenverband} or
\texttt{gesetzliche\ Krankenversicherung}. Some source tables use the
German operational labels \texttt{Kasse} or \texttt{Kassen}; the
manuscript renders these as website entities, insurer entities, or
statutory sickness funds where possible. In those contexts, the unit is
an affected statutory-health-insurance website entity or insurer entity,
not an individual insured person. \texttt{Satzung} denotes an
insurer-specific statutory bylaw or benefit rule that can condition how
statutory benefits are communicated on an individual insurer's website.

Raw crawl data were treated as read-only inputs. KassenWaechter, the
audit workflow used for this study, converted crawl records into
normalized page records with page-level metadata, content hashes, and
review state. The normalized page metadata are the source of truth for
page-level review state; corpus coverage summaries aggregate this
metadata but do not replace the page-level records.

Page eligibility was determined before in-depth review. The crawler
built the candidate URL set from discovered or configured sitemaps,
retained HTML-like German-language pages within configured domains and
path rules, removed obvious static assets and non-German path variants,
and de-duplicated URLs during collection. Main-content extraction
removed common navigation, footer, script, template, cookie-consent, and
auxiliary scaffolding. Content auditability was treated as a
crawler-level quality gate rather than as an audit result: auditable
pages met the configured minimum word-count and URL requirements, while
failed crawl rows, non-HTML assets, non-German path variants, and
private or interactive user data are outside the analytic denominator.

\subsection{Terminology And Evidence
Status}\label{terminology-and-evidence-status}

{\def\LTcaptype{none} 
\begin{longtable}[]{@{}
  >{\raggedright\arraybackslash}p{(\linewidth - 4\tabcolsep) * \real{0.3333}}
  >{\raggedright\arraybackslash}p{(\linewidth - 4\tabcolsep) * \real{0.3333}}
  >{\raggedright\arraybackslash}p{(\linewidth - 4\tabcolsep) * \real{0.3333}}@{}}
\toprule\noalign{}
\begin{minipage}[b]{\linewidth}\raggedright
Term
\end{minipage} & \begin{minipage}[b]{\linewidth}\raggedright
Meaning in this manuscript
\end{minipage} & \begin{minipage}[b]{\linewidth}\raggedright
What it does not mean
\end{minipage} \\
\midrule\noalign{}
\endhead
\bottomrule\noalign{}
\endlastfoot
Website entity & A website, portal, or sub-site in the audit corpus; one
statutory sickness fund can contribute more than one entity & An
individual insured person or necessarily one separate sickness fund \\
Page-level review state & The recorded position of a page in the audit
process, such as in-depth review or lower-priority triage & A legal,
medical, editorial, or quality judgment \\
Review record & A reviewable statement, claim-like passage, or page
segment stored with context, evidence status, materiality fields, and
counterargument fields & A confirmed error, page-level accusation, or
public finding \\
Provenance signal (P) & A signal that may help explain or route
AI-assisted or formulaic content production & Evidence that the content
is wrong \\
Substantive quality signal (Q) & A signal that may warrant medical,
legal, benefit-related, or editorial review & A final determination
without source and human review \\
Public claim & A released page-level or insurer-level assertion &
Something produced by model output alone \\
Case-review queue & The set of review records routed for case-level
human, legal, medical, or editorial review & A list of confirmed
findings \\
Evidence base & The frozen set of stored review records and aggregate
counters underlying the Results tables (dated 2026-05-18) & A set of
adjudicated findings \\
Review axis & The thematic dimension under which a review record or
sampled page is filed, such as benefit, temporal, legal or source,
medical, editorial, or evidence review & An independent clinical or
legal endpoint \\
Flagged review case & A case routed into the paired-reliability subset
through recall-oriented or targeted review routes & A confirmed problem
case \\
\end{longtable}
}

The first analytic distinction is between provenance-oriented signals
(P) and substantive quality signals (Q). P signals help route pages that
may show AI-assisted, formulaic, or mechanically assembled production
patterns. Q signals identify passages that may need medical, legal,
benefit-related, or editorial review. A high P value is not evidence
that a page is wrong, and a low P value is not evidence that a page is
safe. Conversely, a high Q value can occur in human-written or archived
content.

Each page was assigned to a review state through the audit pipeline. A
lower-priority page did not meet the current routing threshold for
in-depth review after deterministic screening and lower-cost triage. An
in-depth-review page went through a deeper review path and has
structured audit metadata. These labels are process states, not legal
conclusions: in-depth review marks a completed review path, and lower
priority marks low routing priority under the current workflow, not a
guarantee that the page is free of issues. At the corpus coverage
snapshot used for this manuscript, all pages had a recorded terminal or
current review state: 18,215 pages were in-depth-review pages, 37,983
were lower-priority pages, no pages remained awaiting in-depth review,
and no pages lacked a recorded state.

Public page-level or insurer-level claims require stricter
public-reporting requirements: a located page quote, preserved source
context, official-source check where applicable, counterargument review,
and documented human adjudication.

\subsection{Audit Workflow And Evidence
Artifacts}\label{audit-workflow-and-evidence-artifacts}

The audit architecture was implemented as a staged workflow in which
each step produced a different type of evidence. This distinction is
important because checked review records are not interchangeable with
raw model outputs.

{\def\LTcaptype{none} 
\begin{longtable}[]{@{}
  >{\raggedright\arraybackslash}p{(\linewidth - 6\tabcolsep) * \real{0.2500}}
  >{\raggedright\arraybackslash}p{(\linewidth - 6\tabcolsep) * \real{0.2500}}
  >{\raggedright\arraybackslash}p{(\linewidth - 6\tabcolsep) * \real{0.2500}}
  >{\raggedright\arraybackslash}p{(\linewidth - 6\tabcolsep) * \real{0.2500}}@{}}
\toprule\noalign{}
\begin{minipage}[b]{\linewidth}\raggedright
Stage
\end{minipage} & \begin{minipage}[b]{\linewidth}\raggedright
Implementation role
\end{minipage} & \begin{minipage}[b]{\linewidth}\raggedright
Primary output
\end{minipage} & \begin{minipage}[b]{\linewidth}\raggedright
Evidentiary status in this manuscript
\end{minipage} \\
\midrule\noalign{}
\endhead
\bottomrule\noalign{}
\endlastfoot
Crawl-to-record compile & Read-only crawl records are normalized into
page records with metadata & Page-level normalized records & Operational
corpus and page metadata \\
Deterministic detector & Configured screening rules assign P and Q
signals and recall routes & Page-level routing metadata & Screening and
prioritization signal \\
Lower-cost triage & LLM prefilter separates lower-priority pages from
in-depth review candidates & Page-level review state & Process state,
not final quality judgment \\
In-depth review & Domain-grounded LLM review emits structured review
records & Structured audit metadata & Candidate evidence before minimum
evidence checks \\
Minimum evidence checks & Code-level filters remove or downgrade
unsupported review records & Checked review records and aggregate
counters & Minimum evidence requirement for aggregate inclusion \\
Aggregate summaries and recurring patterns & Derived summaries group
review records and cross-insurer concepts & Redacted aggregate summaries
and pattern tables & Analysis layer for Results tables \\
Human adjudication with specialist review where required & Case-level
review checks quote, context, legal or medical source, and
counterargument & Stable adjudication record & Required before public
page-level or insurer-level claims \\
\end{longtable}
}

Figure 1 summarizes this architecture as an evidence-boundary schematic.
It should be read as an evidence ladder: broad page routing comes first,
model-generated review evidence is filtered next, aggregate results are
reported only as workload and pattern summaries, and public page-level
claims require documented human adjudication, with medical or legal
specialist review where required.

\begin{figure}[htbp]
\centering
\includegraphics[width=\linewidth]{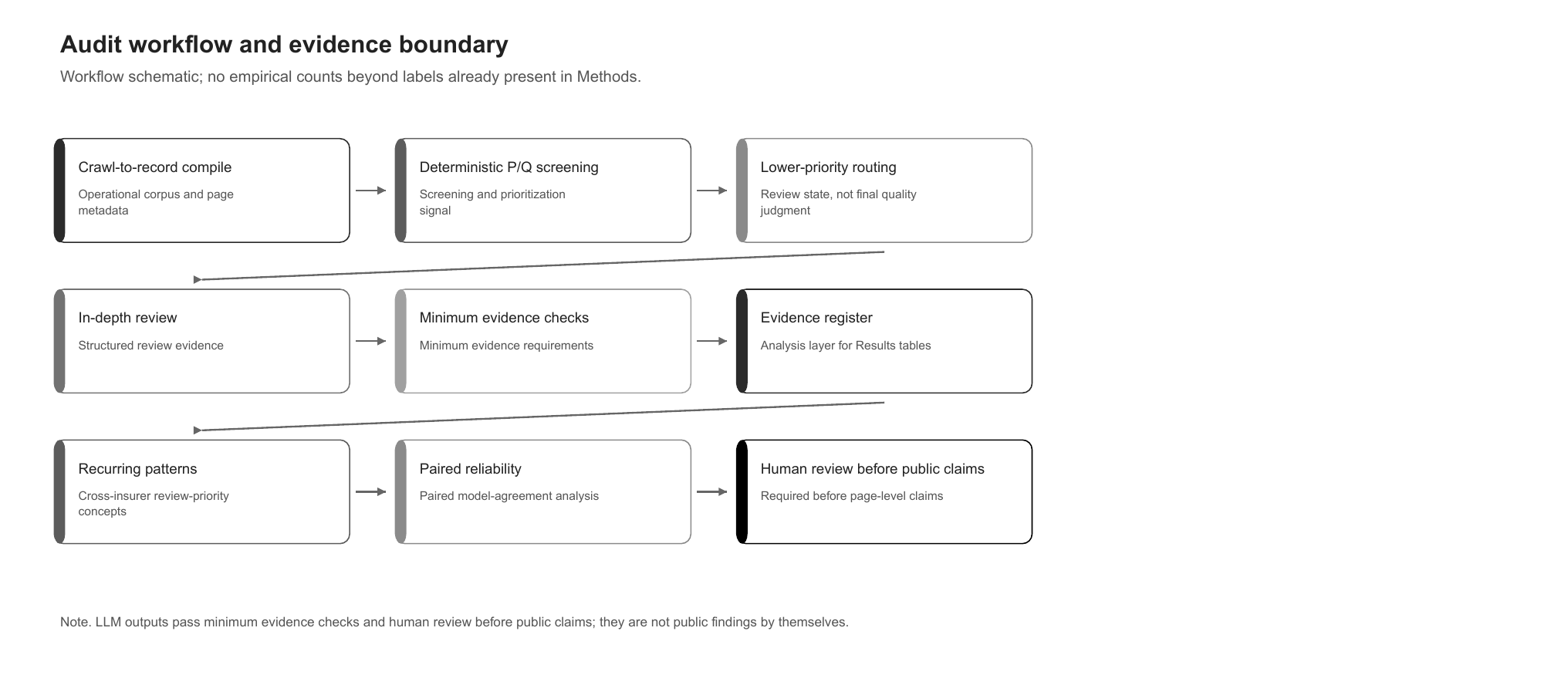}
\caption{Audit workflow and evidence boundary. Schematic overview of page compilation, P/Q screening, lower-cost triage, in-depth review, page-evidence checks, review-record grouping, recurring-pattern analysis, paired reliability analysis, and human review before public claims. The schematic distinguishes page-level review states, model-generated review evidence, checked record outputs, aggregate analysis layers, and human adjudication boundaries, with medical or legal specialist review where required.}
\label{fig:workflow-evidence-boundary}
\end{figure}

Pages selected by deterministic routing, metadata rules, or risk
heuristics were submitted to a lower-cost LLM prefilter. The prefilter
separated lower-priority pages from pages awaiting in-depth review.
In-depth review was then applied to pages routed from deterministic
screening, lower-cost triage, medical or legal context rules,
recall-oriented samples, or targeted case-review workflows. The review
prompt produced structured review records rather than free-form
commentary.

In-depth review incorporated domain context through selective prompt
augmentation. HWG context was available for health-related advertising
or source-check candidates, and SGB V or statutory bylaw (Satzung)
context could be added when a page concerned statutory benefits,
contributions, eligibility, reimbursement, or comparable
public-insurance topics. The reviewed web page remained the evidentiary
object: any page quote used in a review record had to come from the
reviewed page, while external legal or policy context served only to
frame review.

Temporal validity was addressed through selected legal or medical
updates that could postdate model training or page publication. This
directly targets the ``double blindness'' failure mode in which both
page and model may share an outdated view of law or guidance. For
example, a page statement and a reviewing model can both reflect an
outdated legal or medical state if a relevant change occurred after the
page's last update or after the model's training cutoff; the
temporal-validity layer therefore treats selected post-cutoff updates as
routing triggers for source review rather than relying on model
knowledge alone. The manuscript reports the validity role of this layer
rather than its operational implementation; registry structure, topic
triggers, exclusion logic, prioritization rules, prompt wording, and
data-model details are outside the public release boundary.

All in-depth review records were passed through deterministic minimum
evidence checks before being written as audit evidence. At a methods
level, these checks require page evidence, remove unsupported or
untraceable records, demote records whose legal or source support is
incomplete, and route ambiguous high-materiality cases for further
review. Exact check sequences, downgrade rules, prompt fields, and
routing metadata are treated as proprietary operational safeguards
rather than public reproducibility requirements.

Before cross-insurer aggregation, review records were mapped to
canonical concept identifiers, so that synonymous benefit, topic, or
marketing labels for the same subject did not fragment one shared review
priority into separate patterns. A record that resolved to such an
identifier is described as linked to a concept entry, and the
recurring-pattern layer is built over these normalized concepts rather
than over raw page wording. The canonical concept inventory and its
alias-resolution rules are maintained as proprietary configuration;
records that did not resolve to a concept remained in the derived
aggregate and were excluded from the public pattern table.

\subsection{Outcomes And Analysis
Plan}\label{outcomes-and-analysis-plan}

The descriptive endpoints and their separate denominators are fixed in
the Reader Map. A page can yield zero, one, or several review records,
and patterns are non-disjoint. The measures therefore describe audit
process and prioritization, not final legal or medical accuracy, insurer
performance, user impact, or prevalence.

All analyses are descriptive and were computed from the frozen derived
snapshot (corpus coverage 2026-05-19; evidence base 2026-05-18) rather
than from a probability sample. Counts are reported as integers.
Percentages are calculated as the numerator count divided by the
relevant analytic denominator, multiplied by 100, and are rounded to one
decimal place. Denominators are never pooled across analytic levels.
Pages, review records, recurring patterns, lower-priority stress-test
cases, and paired model-agreement cases each carry their own
denominator.

For the single-model, risk-enriched routing stress test, Wilson 95\%
confidence intervals describe binomial uncertainty within the
intentionally enriched 300-page model-review sample. They are not
full-corpus prevalence intervals and must not be multiplied by the
37,983 lower-priority pages or any other corpus denominator. No
hypothesis tests or inferential comparisons were specified for the
present analysis.

\subsection{Calibration And Reliability
Checks}\label{calibration-and-reliability-checks}

The study distinguishes calibration and reliability checks rather than
treating validation as one score. Each layer answers a narrower question
and leaves a different boundary in place.

{\def\LTcaptype{none} 
\begin{longtable}[]{@{}
  >{\raggedright\arraybackslash}p{(\linewidth - 6\tabcolsep) * \real{0.2308}}
  >{\raggedleft\arraybackslash}p{(\linewidth - 6\tabcolsep) * \real{0.3077}}
  >{\raggedright\arraybackslash}p{(\linewidth - 6\tabcolsep) * \real{0.2308}}
  >{\raggedright\arraybackslash}p{(\linewidth - 6\tabcolsep) * \real{0.2308}}@{}}
\toprule\noalign{}
\begin{minipage}[b]{\linewidth}\raggedright
Calibration layer
\end{minipage} & \begin{minipage}[b]{\linewidth}\raggedleft
Denominator
\end{minipage} & \begin{minipage}[b]{\linewidth}\raggedright
What it supports
\end{minipage} & \begin{minipage}[b]{\linewidth}\raggedright
What it does not support
\end{minipage} \\
\midrule\noalign{}
\endhead
\bottomrule\noalign{}
\endlastfoot
Page-state coverage & 56,198 pages & Every page had a recorded review
state & Equal depth of review on every page \\
Minimum evidence checks & 35,998 review records & Stored review records
met minimum page-evidence requirements & Legal, medical, or editorial
truth \\
Single-model routing stress test & 300 enriched lower-priority pages &
Boundary stress test of lower-priority routing & Full-corpus
missed-routing prevalence \\
Paired model agreement & 182 matched review cases & Model-assisted
triage and disagreement-queue calibration & Correctness of individual
review records \\
\end{longtable}
}

None of these layers is a human reference-standard validation study. The
lower-priority routing stress test examines whether routing away from
in-depth review can be treated as a negative label; it cannot.
Additional 100-case model-review checks (Kimi K2.6 and DeepSeek v4 Pro)
are reported in Supplementary Note S2, not as primary endpoints or
tie-breakers.

The single-model, risk-enriched routing stress test drew 300 pages from
lower-priority pages meeting an 80-word minimum threshold and
intentionally enriched strata more likely to contain residual review
need. One Gemini 3.1 Pro reviewer (Vertex AI) assigned one of three
review-routing labels: no review signal, possible review need, or
material review candidate. Without a human reference standard, the test
cannot estimate sensitivity or determine the direction of routing bias.
Sampling strata and seed are reported in Supplementary Note S2.

Inter-model reliability was evaluated on 191 flagged review cases drawn
from recall-oriented and targeted review routes. The two reviewers were
a DeepSeek v4 Pro configuration with maximum reasoning effort (reviewer
model A) and a Claude Sonnet 4.6 configuration (reviewer model B); model
identities follow the released snapshot metadata. The two model outputs
each contained 191 records; 182 common records could be paired for
analysis, producing a 44-case disagreement queue for adjudication. The
cases were drawn from axes including medical precision, factual dates,
legal-operational statements, contribution values, and trust signals.
The reliability estimate is therefore a paired model-agreement analysis
over the joined subset, not a full human-reference validation set.
Operational labels were not final adjudication labels: retained for
review means retained for the paired review purpose, source-check
required means the case required external legal, official-source,
medical, or specialist review, and low current review value means the
item lacked sufficient case or review value at that stage.

Observed agreement P\_o was computed as the proportion of the n matched
cases in which both audit models assigned the same label.
Chance-expected agreement P\_e was computed from the empirical marginal
label distributions, and Cohen's kappa was then computed as (P\_o -
P\_e) / (1 - P\_e). The resulting values were P\_o = 0.758, P\_e =
0.484, and kappa = 0.532 (asymptotic 95\% confidence interval
0.415-0.649; \citealp{fleiss1969}) for the 182 matched cases. Unweighted
Cohen's kappa is a categorical agreement summary for nominal routing
labels, not a correctness measure or final reliability model. The
conventional \texttt{moderate} descriptor is retained only as a
heuristic, not as validation. Missing-pair arithmetic is reported in
Supplementary Note S2.

The interpretation rule for the manuscript is: counts describe workload;
agreement describes calibration; lower-priority sample yield describes a
boundary stress test; public claims require human-adjudicated cases.

\subsection{Exploratory Public Observability
Coding}\label{exploratory-public-observability-coding}

Exploratory public-observability coding assessed whether predefined
content-governance signals were discoverable across the same 84 website
entities under a defined search path.

The policy-core subset coded five signals: visible editorial standards,
visible content responsibility, content-correction pathways,
update-cycle statements, and AI-related content disclosure. Four broader
or noisier signals, namely page-level review metadata, source-standard
visibility, public AI policy, and cross-insurer content standards, were
excluded from the present analysis. For each entity, the search path
covered the homepage, imprint, privacy notice, contact or feedback
pages, about or transparency pages, health-information or magazine
sections, sitemap or site search, and documented domain-restricted
external search queries where needed.

Coding categories were \texttt{present\_public},
\texttt{partial\_public}, \texttt{not\_found\_in\_path},
\texttt{not\_applicable}, and \texttt{needs\_manual\_review}. The
project-internal policy-core artifact contains 420 cells (84 entities ×
5 signals); it is not part of the public data package. Its first coding
is a model-assisted pass identified as
\texttt{codex\_live\_first\_pass}; it followed the defined search path
and recorded a code, URL, quote, and claim boundary for each cell. The
artifact does not permit reconstruction of the extent of human
involvement. A project-internal, non-independent consistency check
confirmed 419 cells and changed one; the artifact does not identify the
second reviewer's identity, date, or role.

\subsection{Reproducibility
Boundaries}\label{reproducibility-boundaries}

The archived derived snapshot is the reproducibility object; Data
Availability gives the support matrix and separates public recomputation
from proprietary or provider-dependent steps. The boundary preserves
page-state, aggregate, pattern-row, and paired-matrix evidence, while
pattern configuration, prompt assets, provider metadata, raw page text,
and changing external APIs remain outside public re-execution. The final
bibliographic and legal sources define the external reference set.

\section{Results}\label{results}

\subsection{Research Question 1 (RQ1): Prioritized Review
Workload}\label{research-question-1-rq1-prioritized-review-workload}

The first research question asks what review workload remained after
page-evidence checks. The workflow generated the record, concept, and
pattern counts defined in the Reader Map; 21,452 records were
additionally routed into the case-review queue. Counts at different
levels must not be divided into one another.

\subsubsection{Recurring Review
Priorities}\label{recurring-review-priorities}

The central descriptive result of RQ1 is the recurring-pattern layer.
The cross-insurer pattern analysis grouped review records into 290
recurring review patterns: 42 Tier-1 recurring review priorities
(14.5\%), 31 Tier-2 priorities (10.7\%), and 217 Tier-3 priorities
(74.8\%). These tiers are review-priority tiers, not severity labels.
Tier assignment is rule-based: it combines the number of affected
website entities, the presence of high-severity and material records,
the presence of statutory or health-advertising references, and the
number of entities with located page quotes; exact thresholds are part
of the proprietary configuration. The tier variable is used only for
cross-entity review prioritization; it does not indicate adjudicated
severity or public-report readiness.

The most widely distributed review-candidate pattern concerned HWG- and
advertising-law-related source checks and included 1,874 review records
across 67 website entities, with located page quotes represented in all
67 entities. The largest pattern by review-record count was the
medical-contradiction cluster, with 3,938 records across 24 website
entities. Other patterns concerned legal or source-check candidates,
benefit-promise conditioning checks, benefit-communication transparency,
outdated content, Satzungsleistung or statutory-benefit communication,
and medical claims. The table presents ten Tier-1 patterns selected for
cross-entity breadth and substantive range, not the ten largest by
record count; two omitted Tier-1 patterns contained 1,134 records across
20 entities and 794 across 16. The last column reports the separate
internal routing bucket assigned by the pattern export.

{\def\LTcaptype{none} 
\begin{longtable}[]{@{}
  >{\raggedright\arraybackslash}p{(\linewidth - 8\tabcolsep) * \real{0.1667}}
  >{\raggedleft\arraybackslash}p{(\linewidth - 8\tabcolsep) * \real{0.2222}}
  >{\raggedleft\arraybackslash}p{(\linewidth - 8\tabcolsep) * \real{0.2222}}
  >{\raggedleft\arraybackslash}p{(\linewidth - 8\tabcolsep) * \real{0.2222}}
  >{\raggedright\arraybackslash}p{(\linewidth - 8\tabcolsep) * \real{0.1667}}@{}}
\toprule\noalign{}
\begin{minipage}[b]{\linewidth}\raggedright
Recurring review pattern
\end{minipage} & \begin{minipage}[b]{\linewidth}\raggedleft
Review records
\end{minipage} & \begin{minipage}[b]{\linewidth}\raggedleft
Website entities
\end{minipage} & \begin{minipage}[b]{\linewidth}\raggedleft
Entities with located page quotes
\end{minipage} & \begin{minipage}[b]{\linewidth}\raggedright
Internal routing bucket
\end{minipage} \\
\midrule\noalign{}
\endhead
\bottomrule\noalign{}
\endlastfoot
HWG and advertising-law source checks & 1,874 & 67 & 67 & high \\
Legal or source-check candidates & 980 & 50 & 50 & high \\
Benefit-promise conditioning checks & 1,438 & 47 & 47 & high \\
Benefit communication and transparency & 905 & 45 & 44 & high \\
Outdated content & 621 & 42 & 41 & high \\
Satzungsleistungen and statutory benefits & 533 & 40 & 40 & high \\
Medical contradictions & 3,938 & 24 & 24 & critical \\
Medical claims & 571 & 29 & 28 & high \\
Possible medical inaccuracy & 414 & 24 & 24 & high \\
Vaccination-related benefit promises & 211 & 24 & 24 & high \\
\end{longtable}
}

The internal routing bucket is an internal materiality and routing
label, not a severity rating; a critical bucket denotes internal review
priority, not adjudicated severity.

This distribution supports two observations. First, the review need is
not confined to a small number of conspicuous websites; several Tier-1
recurring review priorities recur across dozens of website entities.
Second, the leading clusters are predominantly Q-oriented: they concern
the quality, conditioning, temporal validity, legal framing, or
evidentiary support of published statements. P-oriented provenance
signals remain relevant for explaining possible production pathways and
prioritizing review, but they do not carry the substantive result. This
result supports cross-entity prioritization; it does not support public
legal or medical claims about any single page or insurer.

Concept-linked pattern analysis intentionally reports only the
normalized, recurring layer. Of 33,448 concept-eligible records, 23,273
(69.6\%) linked to concept entries, while 10,175 stayed unmatched and
2,550 of the 35,998 stored records were outside the matching step. The
290 public pattern rows contain 21,459 assignments over 17,076 distinct
records, because one record can appear under several patterns. The
remaining 6,197 concept-linked records did not enter a recurring
pattern, since a pattern required recurrence across entities. Unmatched
and non-eligible records remain part of the workload, but they do not
contribute to the public recurring-pattern table.

\subsubsection{Broad Review Categories}\label{broad-review-categories}

The five largest categories in the 2026-05-18 evidence base were
transparency, legal framing, medical content, contradictions, and
AI-related failure-mode signals. These categories are schema groupings
for review prioritization, not independent clinical or legal endpoints.
Together they accounted for between roughly 88\% and 91\% of records on
every measure -- review records, material records, page-quote-located
records, and records routed into the case-review queue alike -- with the
exact per-category counts in the table below.

{\def\LTcaptype{none} 
\begin{longtable}[]{@{}
  >{\raggedright\arraybackslash}p{(\linewidth - 16\tabcolsep) * \real{0.0857}}
  >{\raggedleft\arraybackslash}p{(\linewidth - 16\tabcolsep) * \real{0.1143}}
  >{\raggedleft\arraybackslash}p{(\linewidth - 16\tabcolsep) * \real{0.1143}}
  >{\raggedleft\arraybackslash}p{(\linewidth - 16\tabcolsep) * \real{0.1143}}
  >{\raggedleft\arraybackslash}p{(\linewidth - 16\tabcolsep) * \real{0.1143}}
  >{\raggedleft\arraybackslash}p{(\linewidth - 16\tabcolsep) * \real{0.1143}}
  >{\raggedleft\arraybackslash}p{(\linewidth - 16\tabcolsep) * \real{0.1143}}
  >{\raggedleft\arraybackslash}p{(\linewidth - 16\tabcolsep) * \real{0.1143}}
  >{\raggedleft\arraybackslash}p{(\linewidth - 16\tabcolsep) * \real{0.1143}}@{}}
\toprule\noalign{}
\begin{minipage}[b]{\linewidth}\raggedright
Broad category
\end{minipage} & \begin{minipage}[b]{\linewidth}\raggedleft
Review records
\end{minipage} & \begin{minipage}[b]{\linewidth}\raggedleft
Share of review records
\end{minipage} & \begin{minipage}[b]{\linewidth}\raggedleft
Material records
\end{minipage} & \begin{minipage}[b]{\linewidth}\raggedleft
Material share within category
\end{minipage} & \begin{minipage}[b]{\linewidth}\raggedleft
Page-quote-located records
\end{minipage} & \begin{minipage}[b]{\linewidth}\raggedleft
Page-quote-located share within category
\end{minipage} & \begin{minipage}[b]{\linewidth}\raggedleft
Routed into case-review queue
\end{minipage} & \begin{minipage}[b]{\linewidth}\raggedleft
Case-review queue share within category
\end{minipage} \\
\midrule\noalign{}
\endhead
\bottomrule\noalign{}
\endlastfoot
Transparency & 11,887 & 33.0\% & 7,603 & 64.0\% & 9,988 & 84.0\% & 6,243
& 52.5\% \\
Legal framing & 8,801 & 24.4\% & 7,740 & 87.9\% & 8,087 & 91.9\% & 7,071
& 80.3\% \\
Medical content & 6,136 & 17.0\% & 4,981 & 81.2\% & 5,531 & 90.1\% &
4,452 & 72.6\% \\
Contradictions & 3,130 & 8.7\% & 1,709 & 54.6\% & 2,122 & 67.8\% & 1,023
& 32.7\% \\
AI-related failure-mode signals & 2,191 & 6.1\% & 423 & 19.3\% & 1,975 &
90.1\% & 393 & 17.9\% \\
\end{longtable}
}

Figure 2 visualizes the same five-category profile, using review-record
counts together with within-category materiality, page-quote-location,
and case-review queue shares.

\begin{figure}[htbp]
\centering
\includegraphics[width=\linewidth]{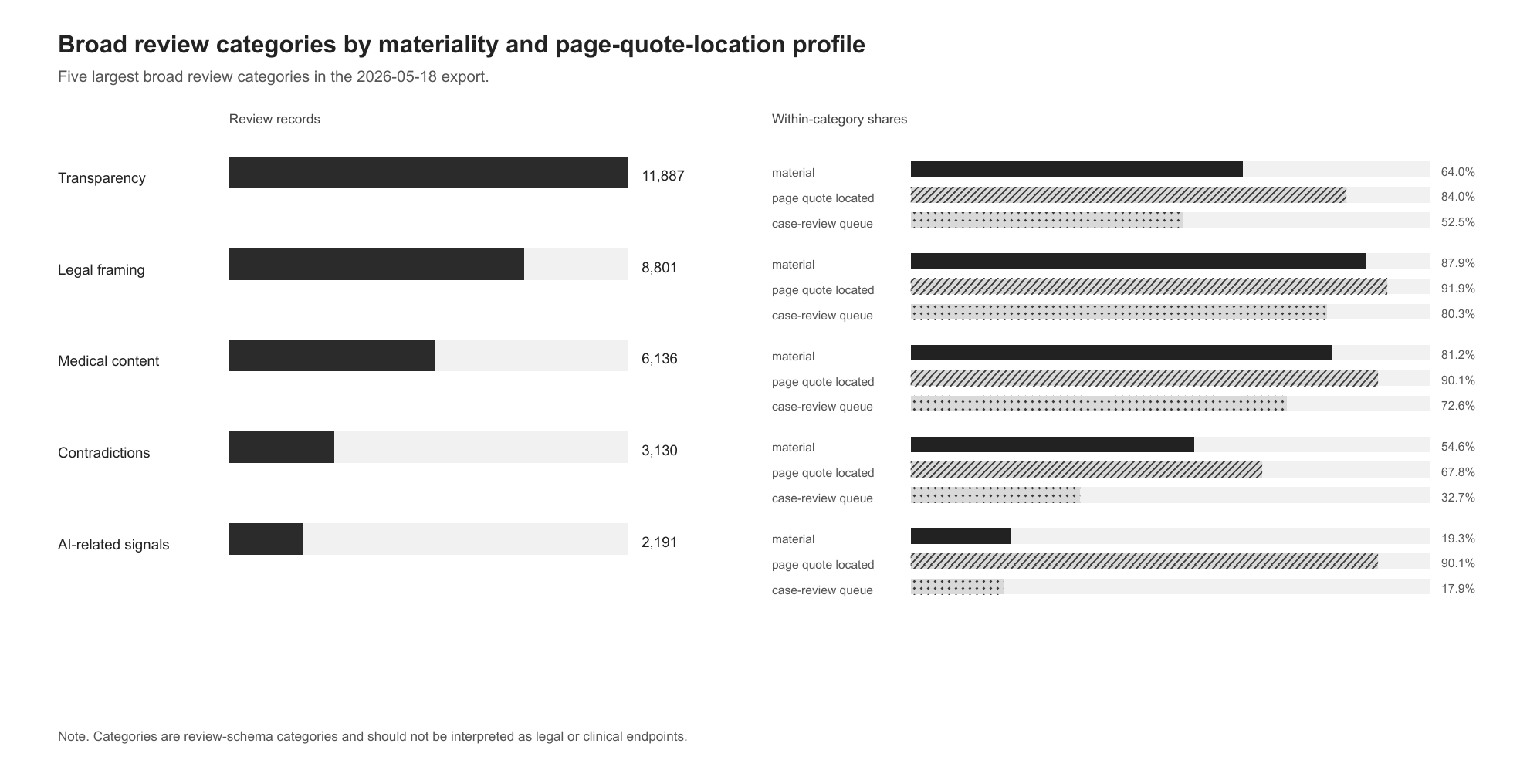}
\caption{Broad review categories by materiality and page-quote-location profile. Horizontal grouped bar chart for the five largest categories, showing review-record counts, material-record shares, page-quote-location shares, and case-review queue shares. Categories are workflow schema categories, not independent legal or clinical endpoints.}
\label{fig:broad-category-profile}
\end{figure}

The displayed labels are English renderings of the original schema codes
\texttt{Transparenz}, \texttt{Recht}, \texttt{Medizin},
\texttt{Widerspruch}, and \texttt{KI-Fail}. These categories show that
the dominant review workload is concentrated in Q-oriented domains:
benefit and transparency communication, legal conditioning,
medical-content review, and contradictions across page text or source
context. The high within-category material and case-review shares for
the legal-framing and medical-content categories indicate that legal and
medical categories were frequent and comparatively enriched for records
retained as material review records. The AI-related failure-mode
category is analytically different. It captures provenance, currentness,
or model-risk signals that can help explain production pathways or
prioritize review, but its lower material and case-review shares show
why it should not be interpreted as direct evidence of a substantive
medical or legal issue.

\subsubsection{Review Workload Profile}\label{review-workload-profile}

As of the 2026-05-18 evidence base, the workflow stored 35,998
model-generated review records. All underwent deterministic minimum
evidence checks and carry the resulting evidence-status category: the
quoted passage was located in the captured page text for 31,347 records,
not located for 4,631, and recorded as missing quote status for 20.
Undergoing the checks is not the same as clearing them; the 4,631
not-located and 20 missing-status records remain in the workload and
require quote or evidence review before any public use. Location
confirms literal occurrence in the captured text, not factual
correctness. Of the records, 24,791 were material, while the remainder
were \texttt{soft} or otherwise non-material observations retained for
traceability. The summary also routed 21,452 records into the
case-review queue. These records are a capacity-planning output
requiring human, legal, medical, or editorial assessment before any
public claim.

The same summary separates substantive-risk records from trust-signal or
provenance-oriented records. Of the 35,998 review records, 32,036 were
assigned to the substantive-risk track and 3,962 to the
trust/provenance-signal track. Suggested review labels were distributed
as 22,462 retained-for-review records, 8,843 records requiring legal,
official-source, medical, or specialist source checking, and 4,693
records treated as low current review value. These are routing labels,
not final human adjudication labels.

{\def\LTcaptype{none} 
\begin{longtable}[]{@{}
  >{\raggedright\arraybackslash}p{(\linewidth - 6\tabcolsep) * \real{0.2143}}
  >{\raggedleft\arraybackslash}p{(\linewidth - 6\tabcolsep) * \real{0.2857}}
  >{\raggedleft\arraybackslash}p{(\linewidth - 6\tabcolsep) * \real{0.2857}}
  >{\raggedright\arraybackslash}p{(\linewidth - 6\tabcolsep) * \real{0.2143}}@{}}
\toprule\noalign{}
\begin{minipage}[b]{\linewidth}\raggedright
Review-record property
\end{minipage} & \begin{minipage}[b]{\linewidth}\raggedleft
Count
\end{minipage} & \begin{minipage}[b]{\linewidth}\raggedleft
Descriptive share
\end{minipage} & \begin{minipage}[b]{\linewidth}\raggedright
Interpretation
\end{minipage} \\
\midrule\noalign{}
\endhead
\bottomrule\noalign{}
\endlastfoot
Review records & 35,998 & 100.0\% & Generated review records in the
2026-05-18 evidence base, not confirmed violations \\
Material review records & 24,791 & 68.9\% & Records not treated as
\texttt{soft} observations \\
Page-quote-located records & 31,347 & 87.1\% & Records whose quoted
passage was located in the reviewed page text \\
Quote-not-located records & 4,631 & 12.9\% & Records requiring quote or
evidence review before public use \\
Missing quote status & 20 & 0.1\% & Records requiring cleanup or
exclusion before final analysis \\
Records routed into case-review queue & 21,452 & 59.6\% &
Capacity-planning queue for case-level review, not final public
claims \\
Substantive-risk track records & 32,036 & 89.0\% & Records assigned to
substantive risk review \\
Trust/provenance-signal track records & 3,962 & 11.0\% & Records
assigned to provenance or trust-signal review \\
\end{longtable}
}

Figure 3 presents these aggregate quantities as overlapping record
properties rather than as a funnel, preserving the distinction between
generated review records, materiality, page-quote-location, case-review
routing, and track assignment.

\begin{figure}[htbp]
\centering
\includegraphics[width=\linewidth]{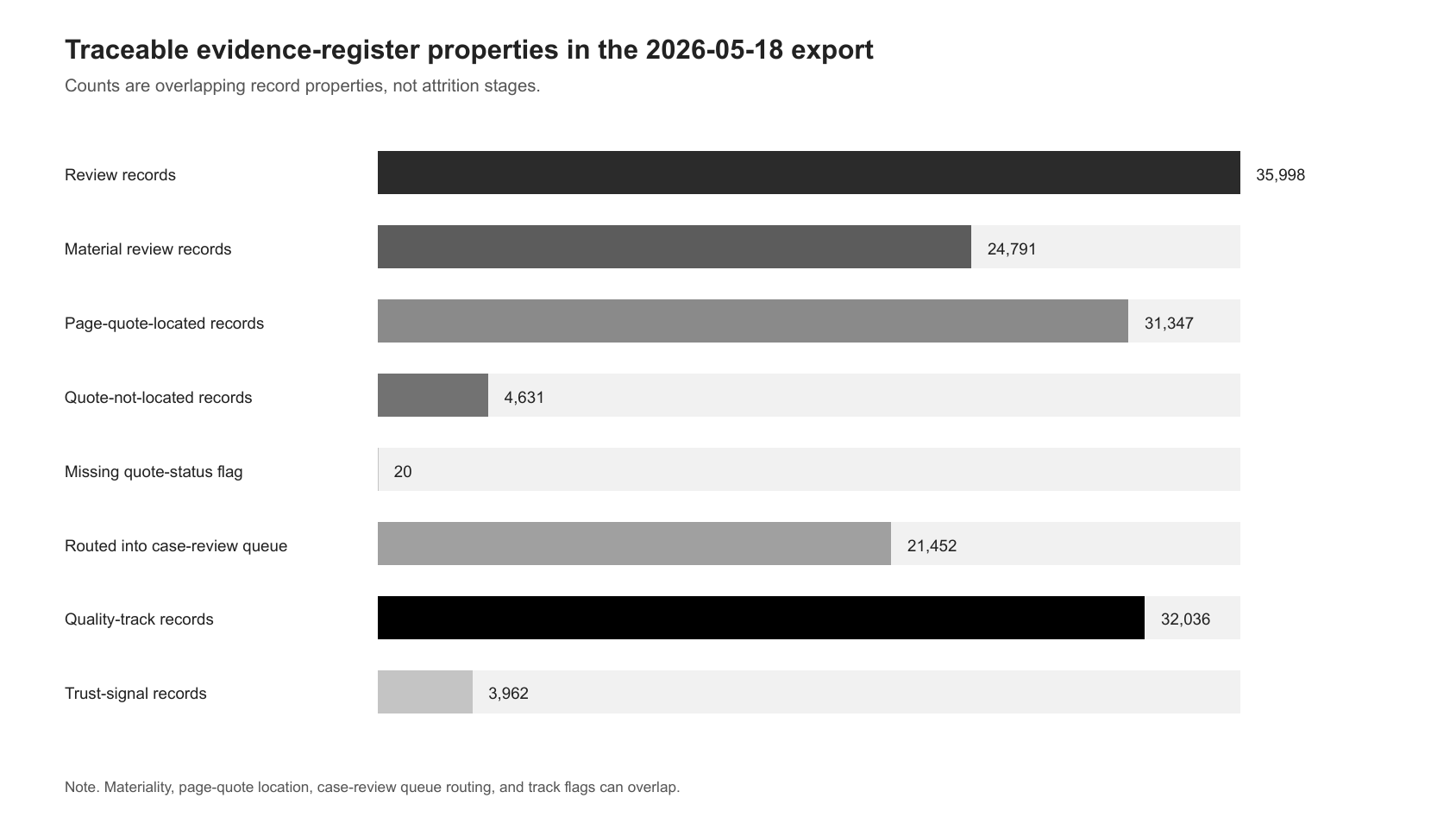}
\caption{Review-record properties in the 2026-05-18 evidence base. Grouped bar chart of review records, material review records, page-quote-located records, quote-not-located records, missing quote-status records, records routed into the case-review queue, substantive-risk track records, and trust/provenance-signal track records. Counts are overlapping record properties and should not be interpreted as mutually exclusive attrition stages.}
\label{fig:evidence-register-profile}
\end{figure}

Overall, RQ1 supports the claim that the workflow produces a structured
review workload at corpus scale. It does not support prevalence
estimates, insurer rankings, or final determinations about legal or
medical correctness.

\subsection{Research Question 2 (RQ2): Corpus Coverage and
Lower-Priority
Routing}\label{research-question-2-rq2-corpus-coverage-and-lower-priority-routing}

The second research question asks whether the workflow can assign a
current review state to the complete corpus while preserving the
distinction between broad triage and in-depth review. Coverage is
reported here as a validity condition for interpreting the review
workload: the denominator is known, but the counts above remain outputs
of risk-based routing and minimum-evidence review rather than prevalence
estimates.

\subsubsection{Corpus Completion And Page-Level Review
States}\label{corpus-completion-and-page-level-review-states}

The corpus coverage snapshot used for this manuscript shows complete
page-state coverage for the current corpus. All 56,198 pages had a
recorded state; 18,215 pages (32.4\%) were in the in-depth review state
and 37,983 pages (67.6\%) were in the lower-priority triage state. No
pages remained awaiting in-depth review, and no pages lacked a recorded
state. This result indicates operational closure of the audit backlog;
it does not mean that every page received the same depth of review. The
state distribution reflects the intended design: broad deterministic and
lower-cost triage across the corpus, with in-depth review reserved for
pages that crossed routing thresholds or entered targeted review paths.

Figure 4 visualizes this page-state distribution and retains the
zero-count backlog categories to make the closure claim and its
page-state semantics explicit.

\begin{figure}[htbp]
\centering
\includegraphics[width=\linewidth]{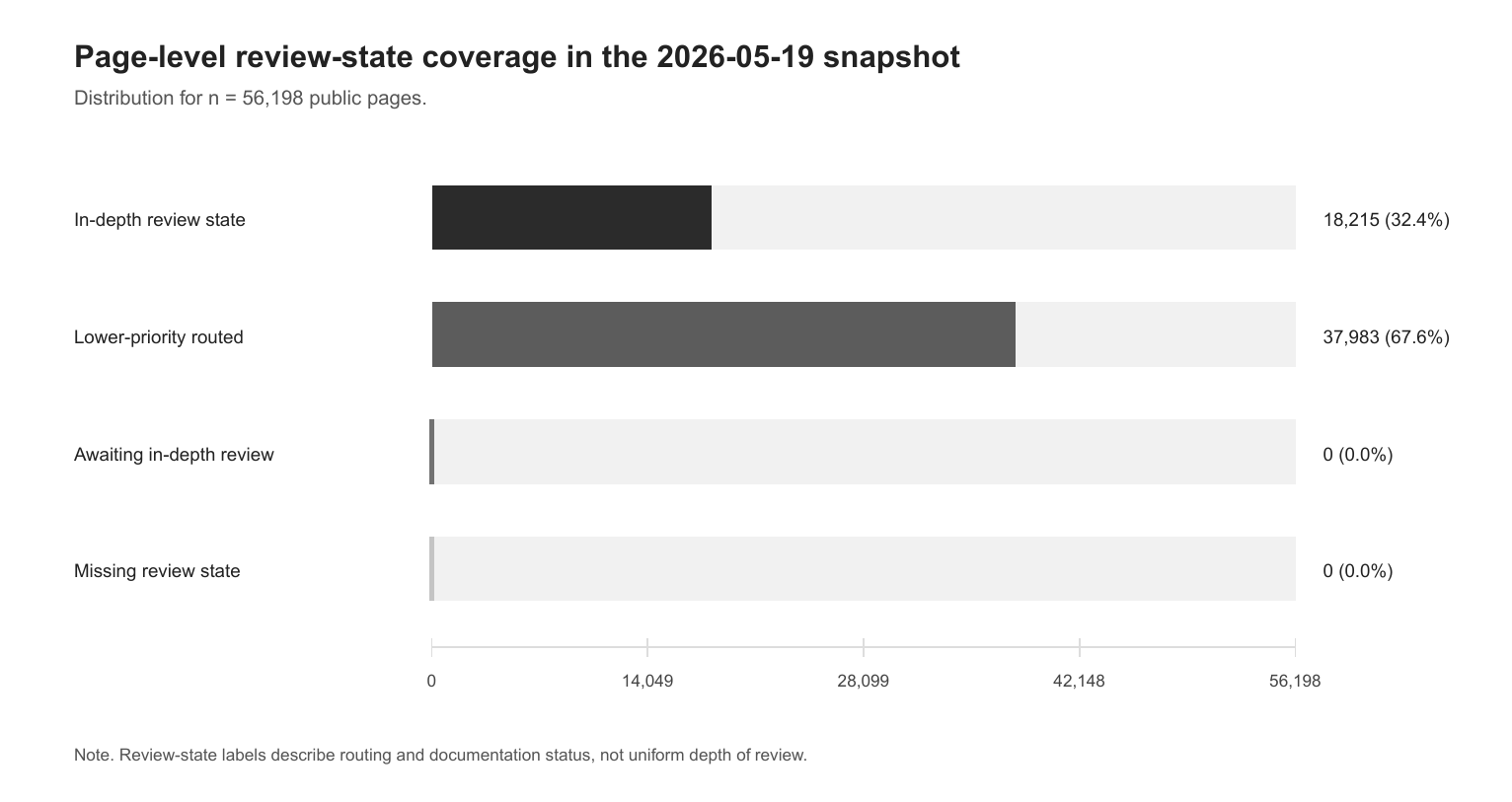}
\caption{Page-level review-state coverage in the 2026-05-19 corpus coverage snapshot. Bar chart showing the distribution for n = 56,198 pages: 18,215 pages in the in-depth review state, 37,983 pages in the lower-priority triage state, 0 pages awaiting in-depth review, and 0 pages without a recorded state. The in-depth review label is a process state and does not imply uniform depth of review.}
\label{fig:page-state-coverage}
\end{figure}

The corpus coverage snapshot also carries one quality-assurance caveat:
a single lower-cost triage run returned all collected results but logged
20 collection errors. This is a provenance note about the frozen
evidence base, not an open audit backlog; the page-level status report
still records every page in a state, with none awaiting in-depth review.
These 20 logged collection errors did not change the page-state
denominator or create unassigned pages in the corpus coverage table.

\subsubsection{Lower-Priority Routing
Checks}\label{lower-priority-routing-checks}

The single-model, risk-enriched routing stress test produced 300 valid
rows with no parser or schema errors: 200 pages had no further review
signal, 58 had possible review need, and 42 were material review
candidates. The model-labelled residual yield was 100/300 (33.3\%;
Wilson 95\% CI 28.2\%-38.8\% within the sample only); the material yield
was 42/300 (14.0\%; Wilson 95\% CI 10.5\%-18.4\%). These outputs are not
sensitivity or full-corpus prevalence estimates.

{\def\LTcaptype{none} 
\begin{longtable}[]{@{}
  >{\raggedright\arraybackslash}p{(\linewidth - 6\tabcolsep) * \real{0.2000}}
  >{\raggedleft\arraybackslash}p{(\linewidth - 6\tabcolsep) * \real{0.2667}}
  >{\raggedleft\arraybackslash}p{(\linewidth - 6\tabcolsep) * \real{0.2667}}
  >{\raggedleft\arraybackslash}p{(\linewidth - 6\tabcolsep) * \real{0.2667}}@{}}
\toprule\noalign{}
\begin{minipage}[b]{\linewidth}\raggedright
Stress-test stratum
\end{minipage} & \begin{minipage}[b]{\linewidth}\raggedleft
No review signal
\end{minipage} & \begin{minipage}[b]{\linewidth}\raggedleft
Possible review need
\end{minipage} & \begin{minipage}[b]{\linewidth}\raggedleft
Material review candidate
\end{minipage} \\
\midrule\noalign{}
\endhead
\bottomrule\noalign{}
\endlastfoot
Benefit or legal content & 43 & 23 & 9 \\
General low signal & 29 & 1 & 0 \\
High-signal medical/legal & 59 & 18 & 13 \\
Medical context & 38 & 7 & 15 \\
Style or quality near threshold & 31 & 9 & 5 \\
Total & 200 & 58 & 42 \\
\end{longtable}
}

The most frequent review axes were benefit (34), temporal (25), legal or
source (24), editorial (11), medical (5), and evidence review (1); 200
pages had no review axis. Recommended next steps were discard for 200
pages, source check for 44, human review for 31, specialist review for
14, and in-depth review for 11. The reported run used the prompt dated
2026-05-21; it superseded an earlier undated version with more
aggressive temporal labels.

\subsection{Research Question 3 (RQ3): Calibration Limits from Paired
Model
Review}\label{research-question-3-rq3-calibration-limits-from-paired-model-review}

The third research question reports paired-model consistency and the
adjudication boundary.

\subsubsection{Paired Model Agreement}\label{paired-model-agreement}

The two models produced 191 rows each. Of 182 matched cases (95.3\%), 44
disagreed (24.2\%); nine records per output were unmatched. The paired
statistics were P\_o = 0.758, P\_e = 0.484, and kappa = 0.532.

{\def\LTcaptype{none} 
\begin{longtable}[]{@{}
  >{\raggedright\arraybackslash}p{(\linewidth - 8\tabcolsep) * \real{0.1579}}
  >{\raggedleft\arraybackslash}p{(\linewidth - 8\tabcolsep) * \real{0.2105}}
  >{\raggedleft\arraybackslash}p{(\linewidth - 8\tabcolsep) * \real{0.2105}}
  >{\raggedleft\arraybackslash}p{(\linewidth - 8\tabcolsep) * \real{0.2105}}
  >{\raggedleft\arraybackslash}p{(\linewidth - 8\tabcolsep) * \real{0.2105}}@{}}
\toprule\noalign{}
\begin{minipage}[b]{\linewidth}\raggedright
Reviewer model A label
\end{minipage} & \begin{minipage}[b]{\linewidth}\raggedleft
Reviewer model B: retained for review
\end{minipage} & \begin{minipage}[b]{\linewidth}\raggedleft
Reviewer model B: source-check required
\end{minipage} & \begin{minipage}[b]{\linewidth}\raggedleft
Reviewer model B: low current review value
\end{minipage} & \begin{minipage}[b]{\linewidth}\raggedleft
Total
\end{minipage} \\
\midrule\noalign{}
\endhead
\bottomrule\noalign{}
\endlastfoot
Retained for review & 1 & 0 & 0 & 1 \\
Source-check required & 2 & 57 & 15 & 74 \\
Low current review value & 5 & 22 & 80 & 107 \\
\textbf{Total} & \textbf{8} & \textbf{79} & \textbf{95} &
\textbf{182} \\
\end{longtable}
}

Figure 5 displays this paired-label matrix as a heatmap for the 182
joined records and should be read together with the operational label
definitions in the Methods section.

\begin{figure}[htbp]
\centering
\includegraphics[width=\linewidth]{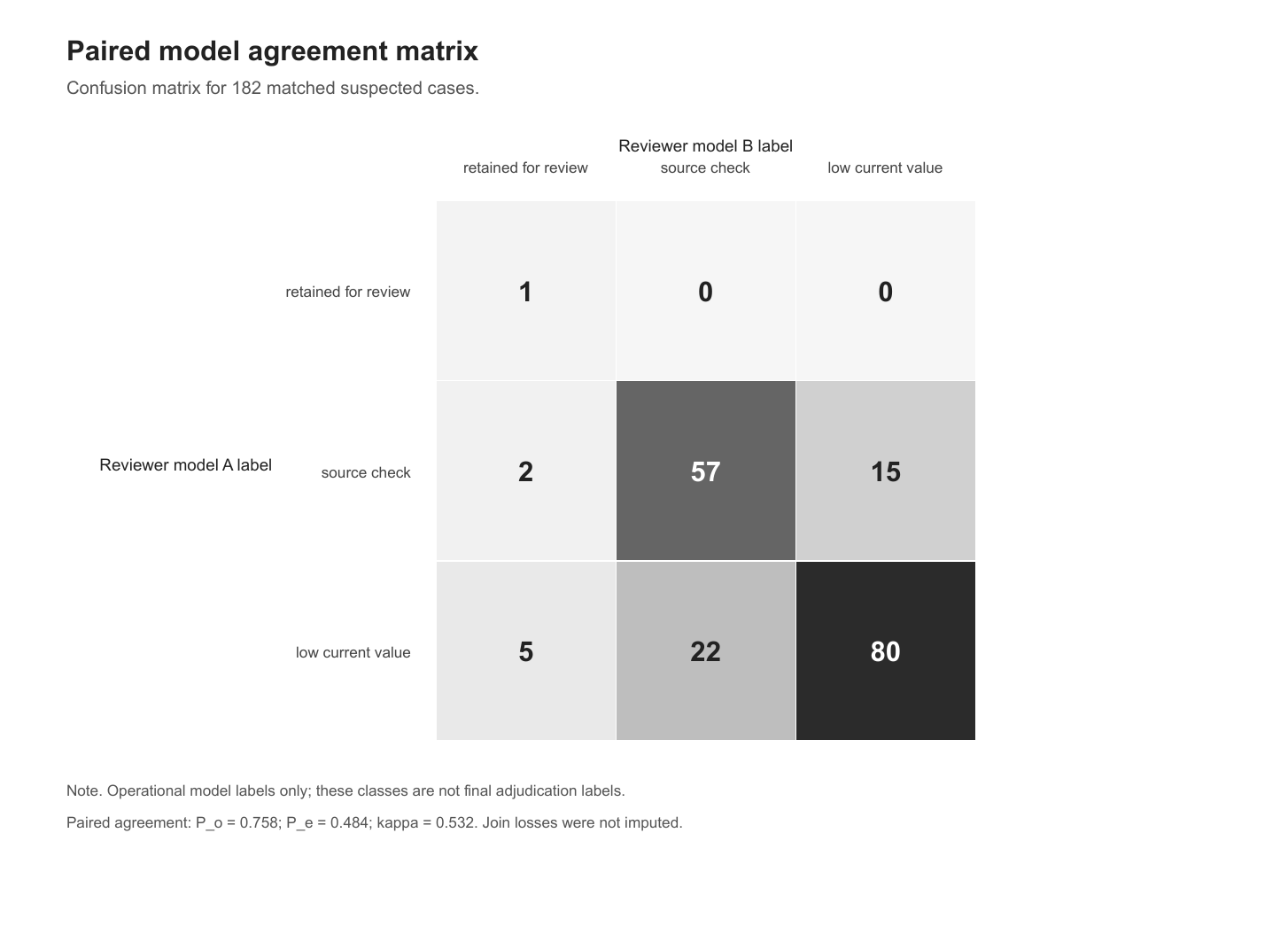}
\caption{Paired model agreement matrix. Confusion-matrix heatmap for 182 matched cases, with Reviewer model A labels as rows and Reviewer model B labels as columns. The paired analysis yielded P\_o = 0.758, P\_e = 0.484, and kappa = 0.532 (95\% CI 0.415-0.649). Retained-for-review, source-check-required, and low-current-review-value labels are operational model-triage labels, not final adjudication labels.}
\label{fig:stage-b-agreement}
\end{figure}

\subsubsection{Human Review Boundary}\label{human-review-boundary}

An internal, model-assisted follow-up (Claude Opus 4.6) reviewed the 8
cases retained by the second model against audit-record context, page
type, quote, legal or medical reference point, and materiality. It
retained 5 cases, marked 2 borderline pending source checks, and 1 false
positive. The record is tied to the dated reliability run, but the
adjudicating model shares a provider family with reviewer model B and is
not an independent human reference standard. Public use still requires
stable case IDs, preserved page context, official-source verification
where applicable, counterargument review, and documented human
adjudication.

\subsection{Secondary Public Observability
Finding}\label{secondary-public-observability-finding}

The exploratory module recorded clear public signals for editorial
standards in 3 entities, content responsibility in 11, correction
pathways in 6, update-cycle statements in 4, and AI disclosure in 3.
Including partial signals, the counts were 15, 54, 32, 4, and 4.

{\def\LTcaptype{none} 
\begin{longtable}[]{@{}
  >{\raggedright\arraybackslash}p{(\linewidth - 8\tabcolsep) * \real{0.1667}}
  >{\raggedleft\arraybackslash}p{(\linewidth - 8\tabcolsep) * \real{0.2222}}
  >{\raggedleft\arraybackslash}p{(\linewidth - 8\tabcolsep) * \real{0.2222}}
  >{\raggedleft\arraybackslash}p{(\linewidth - 8\tabcolsep) * \real{0.2222}}
  >{\raggedright\arraybackslash}p{(\linewidth - 8\tabcolsep) * \real{0.1667}}@{}}
\toprule\noalign{}
\begin{minipage}[b]{\linewidth}\raggedright
Signal
\end{minipage} & \begin{minipage}[b]{\linewidth}\raggedleft
Clear public signal
\end{minipage} & \begin{minipage}[b]{\linewidth}\raggedleft
Partial public signal
\end{minipage} & \begin{minipage}[b]{\linewidth}\raggedleft
Not found in defined public search path
\end{minipage} & \begin{minipage}[b]{\linewidth}\raggedright
Interpretation boundary
\end{minipage} \\
\midrule\noalign{}
\endhead
\bottomrule\noalign{}
\endlastfoot
Editorial standards & 3 & 12 & 69 & Public visibility of editorial
principles or adjacent editorial-team signals, not internal editorial
practice \\
Content responsibility & 11 & 43 & 30 & Explicit author, expert,
quality-review, editorial-team, or general content-responsibility
signals; legal-imprint signals usually counted only as partial \\
Correction pathway & 6 & 26 & 52 & Visible route for content or
editorial feedback; generic service, accessibility, app, seminar, ePA,
and treatment-correction routes were not counted as full
content-correction channels \\
Update cycle & 4 & 0 & 80 & General website or content
review/update-cycle statements, not isolated page dates or
third-party/ePA update notices \\
AI disclosure & 3 & 1 & 80 & Disclosure of AI-generated or AI-assisted
own content or media; general articles about AI or third-party
AI-enabled services were not counted \\
\end{longtable}
}

These counts measure discoverability under the defined path and nothing
beyond it. Absence of a public signal means the signal was not found on
that path; it does not show that the entity lacks an internal editorial
standard, update cycle, or correction process, and a public signal does
not establish that the underlying process is effective.

\section{Discussion}\label{discussion}

\subsection{Principal Findings}\label{principal-findings}

This study shows that public website portfolios of German statutory
health insurance funds can be processed into a structured review
workload at corpus scale while keeping provenance, substantive quality
review, and final adjudication analytically separate. Four results carry
this conclusion: complete page-state coverage of the 56,198-page corpus;
35,998 review records that underwent deterministic minimum evidence
checks, including 31,347 whose quoted passage was located in the
captured page text; 290 non-disjoint recurring patterns with 21,459
record-to-pattern assignments over 17,076 distinct records; and paired
model agreement of 75.8\% (kappa = 0.532) on matched review cases. Quote
location confirms literal occurrence, not factual correctness, and the
outputs describe workload rather than prevalence.

The result is best understood through an implementation-science lens:
the study does not propose a new LLM model, but tests whether a
governed, evidence-preserving audit workflow can operate under real
regulatory, organizational, and source-access constraints. Its
contribution is the operational translation of compliance review into a
scalable applied-data-science process.

The search-path result converges with Scherenberg and Preuss's finding
of limited central discoverability and fragmented provision; greater
provider, editorial, and funding transparency was their recommendation
rather than a measured prevalence finding \citep{scherenberg2023}.

The main methodological finding is that P and Q signals must not be
collapsed. AI-provenance indicators can help explain production
pathways, identify text that may have been assembled or rewritten under
automated workflows, and prioritize review when combined with other
signals. They do not establish that a page is medically inaccurate,
legally insufficiently conditioned, or editorially unsafe. A page can
also contain a reviewable medical, legal, or benefit-related claim when
no strong AI-provenance signal is present. The publishable scientific
object is a review-prioritization method for public health information,
not an AI-detection leaderboard. It is equally distinct from the surface
dimensions that deterministic tools already measure well: readability of
the German health web, for instance, can be scored at scale without
model assistance \citep{zowalla2023}, yet such scores cannot indicate
whether a benefit promise is sufficiently conditioned, a medical
statement is current, or a legal reference is adequate.

Pattern breadth and record volume answer different planning questions.
The HWG-related pattern is the most widely distributed and therefore
identifies a cross-entity review route, whereas the
medical-contradiction pattern is largest by record volume and therefore
represents a different capacity burden. Because records can appear under
several patterns, assignment totals support routing design but cannot be
summed as distinct cases or interpreted as an attrition funnel.

\subsection{Interpretation Of Model
Agreement}\label{interpretation-of-model-agreement}

The paired model comparison quantifies consistency and the expected
divergence workload of two model configurations; it does not establish
correctness or sufficient triage performance. An \texttt{adequate}
result would require a performance or capacity criterion, which was not
measured. Two language models are also not statistically independent
raters: shared training data can produce correlated errors, so agreement
may overstate rather than understate reliability
\citep{zheng2023,chen2025,bedi2025}.

The comparison therefore supports only a bounded workflow
interpretation: models can construct a disagreement queue, while
medical, benefit, Satzung, and legal cases still require evidence checks
and adjudication. Deterministic checks make the intermediate record more
auditable but do not validate the model label.

The paired matrix and stress test answer different questions. The matrix
describes how often two configurations diverge on a flagged subset; the
stress test shows that a lower-priority routing label cannot be treated
as a negative label. Neither supplies a human-reference error rate. The
disagreement queue is therefore an operational workload, while
official-source verification and specialist judgment remain the decision
layer for high-materiality cases.

Minimum evidence checks narrow but do not close this boundary. They test
whether required fields are present, whether a quoted passage occurs
literally in captured page text, and whether unsupported legal or source
claims require downgrade or further review. They cannot determine
whether the quoted statement is factually correct, current, or
adequately conditioned in context.

For workflow design, disagreement has a concrete use even without a
correctness claim. It identifies cases in which one configuration would
route or retain material that the other would not, making the expected
adjudication burden visible. Agreement, by contrast, cannot close a case
because correlated model error, shared source gaps, and common knowledge
cutoffs can survive both reviews. The comparison is therefore a
queue-construction diagnostic rather than a substitute for an acceptance
threshold.

\subsection{Limitations And Threats To
Validity}\label{limitations-and-threats-to-validity}

The corpus is a production website-portfolio audit rather than a random
sample. The denominator is explicit, but review-record counts arise from
deterministic and model-assisted routing decisions and should not be
interpreted as unbiased prevalence estimates for all SHI content or for
the German health web. Page states likewise reflect audit operations,
not uniform depth of review. The single-model, risk-enriched routing
stress test confirms that lower-priority routing is not a human
reference-standard negative label, and in-depth review does not imply
exhaustive review of every possible medical, legal, accessibility, or
UX-writing dimension.

Model outputs remain probabilistic. Minimum evidence checks reduce
weakly supported review records, but they cannot eliminate false
negatives, false positives, extraction errors, or context errors.
Records with quote-not-located or missing quote status, and any
adjudication subset used publicly, still require manual or specialist
review before public claims. The temporal-validity safeguard is
targeted, not exhaustive, and crawler or extraction artifacts may affect
page-level evidence despite page-quote occurrence checks.

Construct validity is limited because the audit measures review need,
not final legal or clinical truth. The broad categories reported as
legal framing, medical content, transparency, contradictions, and
AI-related failure-mode signals are useful for prioritizing evidence
review, but they should not be interpreted as mutually exclusive
diagnoses of page quality. The P/Q split reduces one construct threat by
preventing AI-provenance indicators from being treated as substantive
quality findings; it does not eliminate the need to validate each public
statement against page context and authoritative sources.

The public-observability cells are model-assisted and not independently
human-verified. Before public claims beyond the cautious aggregate
formulation reported here, the rows require independent human
confirmation.

Internal validity is limited by model calibration, routing thresholds,
and evidence extraction. Because none of the calibration layers is a
human reference standard, the workflow cannot support reference-standard
sensitivity or specificity claims without human-adjudicated follow-up.
External validity is limited by corpus construction, jurisdiction,
source hierarchy, legal context, and language. The results generalize
most directly to public SHI website portfolios and to similar
institutional health-information settings.

Operational validity is limited by changing websites, changing law, and
changing model behavior. Legal and medical claims can become outdated
after crawling; model APIs and knowledge cutoffs can change. The frozen
derived snapshot is therefore the submission boundary.

\subsection{Implications And Future
Work}\label{implications-and-future-work}

The immediate research implication is that compliance auditing for
public digital health information should be evaluated as a controlled
workflow, not as a standalone model. The most important next steps are
narrower than ``crawl more pages'': repeat the corpus over time, ground
high-materiality benefit and medical review records in authoritative
sources, publish a stable specialist-adjudicated case set, add a small
human anchor set for lower-priority routing, and test whether returned
review records lead to correction, clarification, rejection, or no
action.

That follow-up would connect routing output to editorial disposition
without turning the present workload into an error count. Each reviewed
record could retain its original evidence status, assigned owner,
source-check outcome, specialist decision where required, and final
disposition. A dated human anchor set could then test how routing
performance changes across models and collection windows. It would also
separate routing stability from downstream editorial value, which the
current design cannot yet estimate. Until such outcomes exist, queue
size describes review capacity demand, and neither retention nor
correction can be inferred from a model label alone.

Public claims should be tiered. Corpus-scale metrics can describe
page-level review-state coverage, review-record volume,
page-quote-location state, recurring-pattern distribution, and model
agreement. Insurer-level or page-level claims require the
public-reporting requirements defined in the Methods. This preserves the
value of large-scale screening while avoiding overinterpretation of
model-generated review records.

The policy implication is that more complete, stable, and openly
documented machine-readable access to authoritative German health
evidence would materially improve content governance. PubMed/NCBI, G-BA,
IQWiG, AWMF, and Cochrane all provide some documented access surface,
but their coverage, licensing conditions, and machine-readable
completeness differ substantially
\citep{ncbiapis,gbarss,gbaais,iqwigresults,iqwigprojects,awmfleitlinien,cochraneapi,cochranelibraryaccess}.
Until such infrastructure is more consistent, scalable systems must
combine broad screening with selective, source-grounded verification and
explicit uncertainty reporting.

This source-access point is an implementation observation, not a
documented comparative analysis: the study did not specify systematic
inclusion criteria or produce a results table comparing the source
systems. It supports only the narrower conclusion that automated
cross-source verification was partial and source-dependent in this
workflow.

\section{Conclusion}\label{conclusion}

This snapshot-based audit assigned all 56,198 pages from 84 website
entities a review state, generated 35,998 review records, linked 23,273
of 33,448 eligible records to concepts, and produced 21,459 assignments
across 290 non-disjoint patterns. The design separates provenance from
substantive review and applies temporal and minimum-evidence safeguards.
Paired-model statistics measure consistency and divergence workload, not
correctness. Public page-level or insurer-level claims therefore still
require preserved context, source verification, counterargument review,
and documented human adjudication. Within those limits, the outputs show
that large institutional health-information portfolios can be
transformed into a traceable review corpus.

Exploratory, model-assisted public-observability coding indicated that
editorial standards, responsibility, correction routes, update logic,
and AI disclosure were often not discoverable under the defined search
path.

The aggregate result requires independent human confirmation before
stronger observability claims or recommendations directed at individual
entities.

\section{Declarations}\label{declarations}

\subsection{Ethics Statement}\label{ethics-statement}

This study analyzed only publicly accessible institutional website
content and did not involve human participants, private insured-person
data, claims data, patient records, user-interaction logs, protected
portal content, or interaction with website visitors. No formal
institutional ethics approval was sought for this preprint; a
journal-specific ethics or waiver determination should be obtained if
required by the target venue.

\subsection{Correspondence}\label{correspondence}

Correspondence concerning this manuscript should be addressed to Martin
Möller, Martin Möller Digital Platform Consulting; ORCID
\url{https://orcid.org/0009-0009-4352-8765};
\href{mailto:info@martin-moeller.biz}{\nolinkurl{info@martin-moeller.biz}};
\url{https://martin-moeller.biz}.

\subsection{Data Availability}\label{data-availability}

The derived research snapshot is archived at Zenodo under DOI
\href{https://doi.org/10.5281/zenodo.20544691}{10.5281/zenodo.20544691},
released under CC BY 4.0, and includes the frozen aggregate live-status
summary and paired-model comparison artifacts.

{\def\LTcaptype{none} 
\begin{longtable}[]{@{}
  >{\raggedright\arraybackslash}p{(\linewidth - 4\tabcolsep) * \real{0.3333}}
  >{\raggedright\arraybackslash}p{(\linewidth - 4\tabcolsep) * \real{0.3333}}
  >{\raggedright\arraybackslash}p{(\linewidth - 4\tabcolsep) * \real{0.3333}}@{}}
\toprule\noalign{}
\begin{minipage}[b]{\linewidth}\raggedright
Support object
\end{minipage} & \begin{minipage}[b]{\linewidth}\raggedright
Publicly recomputable from the package
\end{minipage} & \begin{minipage}[b]{\linewidth}\raggedright
Not reproducible from the public package
\end{minipage} \\
\midrule\noalign{}
\endhead
\bottomrule\noalign{}
\endlastfoot
Corpus and evidence aggregates & Corpus states 56,198/18,215/37,983 and
aggregate evidence-status values & Crawling, routing, record generation,
and quote checking against captured page text \\
Recurring patterns & 290 pattern rows, tier totals, and the 21,459
assignment sum & Concept inventory, alias rules, pattern score, and the
five-category table \\
Model checks and coding & Paired matrix with P\_o, P\_e, and kappa & The
300-page routing stress test with strata and review axes, both secondary
model-review runs, public-observability coding, and case-level
adjudication \\
\end{longtable}
}

Hashes establish artifact identity, not provenance validity. Raw crawl
records, extracted page text, screenshots, archived HTML, and full-page
Markdown are not redistributed as part of the public package because
they reproduce publisher-controlled web content.

\subsection{Code Availability}\label{code-availability}

KassenWaechter production code, prompts, thresholds, provider adapters,
credentials, and private configuration are not publicly released.
Provider-dependent calls are not reproducible in a bitwise sense because
model versions, credentials, rate tiers, and provider-side updates can
change outputs.

\subsection{Author Contributions}\label{author-contributions}

Martin Möller: conceptualization, methodology, software, investigation,
formal analysis, data curation, visualization, writing - original draft,
writing - review and editing, and project administration. The author had
full access to the data used for the manuscript snapshot and takes
responsibility for the integrity of the data and the accuracy of the
analysis.

\subsection{Funding}\label{funding}

The author received no external funding for this work.

\subsection{Conflicts Of Interest}\label{conflicts-of-interest}

The author is self-employed as Martin Möller Digital Platform Consulting
and develops AußenBlick-GKV, a prospective audit and advisory offering
for public SHI web content that this methodological work is intended to
inform. No insurer funded or commissioned this study. No insurer-level
ranking is reported. No public page-level finding is released from
unadjudicated model output. The workflow is explicitly designed to
prevent automated review records from becoming public accusations
without preserved page context, source checks, counterargument review,
and documented human adjudication.

\subsection{AI Assistance And Model
Use}\label{ai-assistance-and-model-use}

Large language models were used in the audit workflow for triage,
in-depth review, cross-model comparison, and selected evidence-check
support. Their outputs were treated as review signals, not as final
legal, medical, editorial, empirical, or submission determinations.
AI-assisted drafting and editing tools were also used in preparing this
manuscript. No AI system is an author, and the author reviewed and takes
responsibility for the final text.

{\def\LTcaptype{none} 
\begin{longtable}[]{@{}
  >{\raggedright\arraybackslash}p{(\linewidth - 4\tabcolsep) * \real{0.3333}}
  >{\raggedright\arraybackslash}p{(\linewidth - 4\tabcolsep) * \real{0.3333}}
  >{\raggedright\arraybackslash}p{(\linewidth - 4\tabcolsep) * \real{0.3333}}@{}}
\toprule\noalign{}
\begin{minipage}[b]{\linewidth}\raggedright
Workflow role
\end{minipage} & \begin{minipage}[b]{\linewidth}\raggedright
Model use
\end{minipage} & \begin{minipage}[b]{\linewidth}\raggedright
Public evidentiary status
\end{minipage} \\
\midrule\noalign{}
\endhead
\bottomrule\noalign{}
\endlastfoot
Lower-cost triage & Routing support and recall-oriented review-state
assignment & Page-state signal only \\
In-depth review & Structured candidate-record generation & Candidate
evidence before checks \\
Cross-model comparison & Calibration and disagreement-queue construction
& Reliability signal only \\
300-page routing stress test & Primary reviewer
\texttt{gemini-3.1-pro-preview} through a preview endpoint &
Single-model stress-test signal, not a human reference standard \\
Public-observability coding & Model-assisted first pass
\texttt{codex\_live\_first\_pass} plus an internal consistency check &
Exploratory aggregate signal, not independent human verification \\
Drafting and editing & Manuscript preparation and review simulation &
Author-reviewed text \\
\end{longtable}
}

The released dataset v1.0.0 \texttt{model\_use\_summary.tsv} lists
neither the public-observability coding nor the primary reviewer for the
300-page stress test; this manuscript table is authoritative for those
uses.

No prompts, model outputs, or disclosure logs are released as
supplementary files with this preprint; provider-dependent prompts and
configuration remain proprietary as stated in Code Availability.

\subsection{Public Communication
Safeguard}\label{public-communication-safeguard}

Public insurer-level or page-level examples should not be presented as
findings unless they meet the public-reporting requirements defined in
Methods.

\section{Supplementary Material}\label{supplementary-material}

\subsection{Supplementary Note S1: Snapshot
Summary}\label{supplementary-note-s1-snapshot-summary}

The snapshot summary covers corpus coverage
(\texttt{2026-05-19T03:40:25.883017+00:00}), review-workload summaries
(\texttt{2026-05-18T11:25:03.139752+00:00}), the recurring-pattern
summary (\texttt{2026-05-18T11:25:14.006497+00:00}), paired
model-agreement materials from \texttt{2026-05-01}, and the internal
adjudication subset. These timestamps differ by a few days because the
frozen evidence base was assembled in sequential generation steps over
one collection window: the review-workload and recurring-pattern
summaries on 2026-05-18, the corpus-coverage snapshot on 2026-05-19, and
the dated lower-priority review prompt on 2026-05-21, with the paired
model-agreement materials drawn from the earlier 2026-05-01 reliability
run. They describe the same frozen corpus rather than separate datasets.
Public release should use derived summaries, redacted aggregates,
hashes, codebook notes, and public-ready adjudication bundles rather
than full page-text redistribution.

Two result blocks are derived from frozen internal artifacts rather than
from the public package. Their derivation is described here, and the
exact source files are recorded in the versioned revision record that
accompanies this manuscript. The five broad review categories, together
with their materiality, page-quote-location, and case-review shares, are
aggregated from the frozen record-level evidence ledger; the five
reported categories are the five largest, and the next category is an
order of magnitude smaller. The routing stress-test detail, meaning the
stratum distribution, the review axes, and the recommended next steps,
is aggregated from the dated reviewer result file for the 300 sampled
pages. Both blocks are reproducible inside the project environment but
not from the published package, as stated in Data Availability.

\subsection{Supplementary Note S2: Routing Stress-Test And
Paired-Reliability
Detail}\label{supplementary-note-s2-routing-stress-test-and-paired-reliability-detail}

The lower-priority routing sample used seed \texttt{20260521} and
oversampled higher-risk strata: high-signal medical/legal pages (90),
benefit or legal content (75), medical-context pages (60),
near-threshold style or quality pages (45), and general low-signal pages
(30). In that sample, 200 pages had no further review signal, 58 had
possible review need, and 42 were material review candidates.

The 100-page secondary model-review checks tested model dependence in an
enriched group of cases, not population yield. Kimi K2.6 produced 71.0\%
raw label agreement with the primary Gemini reviewer and assigned 31
pages to material review candidate. DeepSeek v4 Pro produced 54.0\% raw
label agreement on the same subset and assigned 7 pages to material
review candidate. On the same subset, Kimi classified 31 cases as
material and DeepSeek 7, showing that the stress-test yield depended
strongly on the reviewer model. The DeepSeek configuration shares its
model family with reviewer model A from the paired reliability analysis;
the two analyses use different samples, dates, and tasks. These checks
are not primary endpoints, tie-breakers, or human reference-standard
validation.

For paired model agreement, statistics used only the 182 matched
records. The nine per-output records that appeared in only one model
output (18 in total) were treated as unmatched records, not
confusion-matrix entries. Conservative observed-agreement lower bounds
are 138 agreements over the 200-record union (69.0\%), or 138/191
(72.3\%) if the nominal per-output sample size is used as denominator.
Cohen's kappa is not recomputed because missing labels prevent a valid
marginal-distribution estimate.

An internal, model-assisted follow-up adjudication record (Claude Opus
4.6) dated 2026-05-01 retained 5 cases in the internal robust-retained
category, classified 2 as borderline pending further source checks, and
classified 1 as a false positive. This record is not a released human
reference-standard dataset; public use of any case requires the
public-reporting requirements defined in Methods.

In the recurring-pattern table, 23,273 of 33,448 eligible review records
linked to concept entries and 10,175 did not. The 33,448 eligible
records are the subset of the 35,998 stored review records that were
eligible for concept matching; the remaining 2,550 records were outside
the matching step. The 290 pattern rows contain 21,459 record-to-pattern
assignments covering 17,076 unique records according to the internal
artifact. Because a record can appear under more than one pattern,
assignments exceed covered records. Unmatched and non-eligible records
remain in the derived aggregate, not the public pattern table.

\bibliography{references}

\end{document}